\documentclass[aps,prb,twocolumn,showpacs,eqsecnum]{revtex4}

\usepackage{graphicx}
\usepackage{bm}

\usepackage{hyperref}

\newcommand{\sea}{\searrow}
\newcommand{\nea}{\nearrow}
\newcommand{\swa}{\swarrow}
\newcommand{\nwa}{\nwarrow}

\newcommand{\Hh}{\hat{H}}

\newcommand{\Gc}{\mathcal{G}}

\newcommand{\nb}{{\bf n}}
\newcommand{\tb}{{\bf t}}

\newcommand{\ib}{{\bf i}}
\newcommand{\jb}{{\bf j}}
\newcommand{\ab}{{\bf a}}
\newcommand{\bb}{{\bf b}}
\newcommand{\eb}{{\bf e}}
\newcommand{\hb}{{\bf h}}

\newcommand{\s}{{\bf s}}

\newcommand{\qb}{{\bf q}}
\newcommand{\pb}{{\bf p}}

\newcommand{\rb}{{\bf r}}

\newcommand{\lan}{\langle}
\newcommand{\ran}{\rangle}
\newcommand{\om}{\omega}

\newcommand{\al}{\alpha}
\newcommand{\be}{\beta}
\newcommand{\ga}{\gamma}
\newcommand{\Ga}{\Gamma}
\newcommand{\de}{\delta}
\newcommand{\De}{\Delta}
\newcommand{\la}{\lambda}
\newcommand{\sig}{\sigma}

\newcommand{\eps}{\varepsilon}

\newcommand{\lt}{\left}
\newcommand{\rt}{\right}
\newcommand{\beq}{\begin{equation}}
\newcommand{\eeq}{\end{equation}}
\newcommand{\beqar}{\begin{eqnarray*}}
\newcommand{\eeqar}{\end{eqnarray*}}

\usepackage{amsmath}
\usepackage{amsfonts}
\usepackage{amssymb}

\newcommand{\fath}{\mathbf{h}}

\newcommand{\fatr}{\mathbf{r}}

\newcommand{\fati}{\mathbf{i}}
\newcommand{\fatj}{\mathbf{j}}

\newcommand{\fatq}{\mathbf{q}}
\newcommand{\fatn}{\mathbf{n}}

\newcommand{\fats}{\mathbf{s}}

\newcommand{\fatsigma}{\bm{\sigma}}

\newcommand{\Gr}{\mathcal{G}}

\newcommand{\F}{\mathrm{F}}

\newcommand{\gr}{\mathrm{gr}}

\newcommand{\ii}{\mathrm{i}}

\newcommand{\dt}{\mathrm{d}}

\newcommand{\intr}[1]{\int\dt #1}

\newcommand{\trps}[1]{#1^{\textsf{T}}}

\newcommand{\htrps}[1]{#1^\dagger}

\newcommand{\ztupel}[2]
            {\left(#1,#2\right)}

\newcommand{\average}[1]{\left\langle #1\right\rangle}
\newcommand{\taverage}[1]{\langle #1\rangle}
\newcommand{\inner}[2]{\left( #1\cdot#2\right)}
\newcommand{\cross}[2]{\left[#1\times#2\right]}

\begin{document}


\title{Anomalous Hall effect in granular ferromagnetic metals \\and effects of weak localization}

\author{Hendrik Meier$^1$, Maxim Yu. Kharitonov$^{1,2}$, and Konstantin B. Efetov$^1$}
\affiliation{ $^1$Institut f\"ur Theoretische Physik III, Ruhr-Universit\"at Bochum, 44780 Bochum, Germany \\
 $^{2}$Materials Science Division, Argonne National Laboratory, Argonne, IL 60439, USA}

\date{\today}


\begin{abstract}
We theoretically investigate the anomalous Hall effect in a system
of dense-packed ferromagnetic grains in the metallic regime.
Using the formalism recently developed for the conventional Hall
effect in granular metals, we calculate the residual anomalous
Hall conductivity $\sigma_{xy}$ and resistivity $\rho_{xy}$ and
weak localization corrections to them for both skew-scattering and
side-jump mechanisms. We find that
the scaling relation between $\rho_{xy}$ and the longitudinal
resistivity $\rho_{xx}$ of the array does not hold, regardless of
whether it is satisfied for the specific resistivities of the
grain material or not.
The weak localization corrections, however, are found to be in
agreement with those for homogeneous metals.
We discuss recent experimental data on the anomalous Hall effect
in polycrystalline iron films in view of the obtained results.
\end{abstract}

\maketitle


\section{Introduction}

The anomalous Hall effect (AHE) in ferromagnetic materials has
been attracting the interest of researchers for decades.
The first theoretical explanation~\cite{karplusluttinger} of AHE
was given by Karplus and Luttinger in 1954.
%
They have shown that, in essence,
the anomalous Hall current arises from
the population imbalance of the electron spin states that
 is transferred into the asymmetry in electron motion via
spin-orbit coupling.
Since then, the theory of AHE has undergone further significant developments
(see, e.g., recent reviews~\onlinecite{wolflemuttalib} and
\onlinecite{Sinitsyn} and references therein). The
interpretation~\cite{SN} of AHE in terms of the Berry phase
concept has fueled additional
interest\cite{taguchiea,jungwirthea,fang} to the problem.

One distinguishes between the \emph{intrinsic} and
\emph{extrinsic} AHE. The intrinsic AHE arises in a perfect
periodic lattice 
subject to spin-orbit coupling. It is due to the topological
properties of the  Bloch states  and does not require any
disorder.
On the contrary, the extrinsic AHE is due to the asymmetric
spin-orbit scattering of spin-polarized electrons on the
impurities of the sample. Two mechanisms termed {\em
skew-scattering}~\cite{smit} and {\em side-jump}~\cite{berger} are
responsible for the extrinsic AHE.
They depend differently on the amount of disorder in the sample
and, as a result, for certain type of disorder the anomalous Hall
(AH) resistivity scales linearly ($\rho_{xy} \propto \rho_{xx}$)
with the longitudinal resistivity for the skew-scattering, and
quadratically ($\rho_{xy} \propto \rho_{xx}^2$) for the side-jump
mechanism. These scaling relations were observed
experimentally~\cite{chienwestgate} in homogeneous systems.
At the same time, for some heterostructure
systems~\cite{xiongetal,xuetal}
considerable deviations from
this scaling law were reported.

At sufficiently low temperatures, the physics of AHE is enriched
by the quantum effects of Coulomb interactions and weak
localization. The Coulomb interaction correction to the AH
conductivity has been shown to vanish
for both skew-scattering and side-jump
mechanisms~\cite{langenfeldwolfle,MW}. Weak localization (WL)
effects were studied in
Refs.~\onlinecite{langenfeldwolfle,dugaevea,MW} and it was
demonstrated that WL correction to AH conductivity is
nonzero\cite{langenfeldwolfle,dugaevea,MW} for skew-scattering and
vanishes~\cite{dugaevea,MW} for side-jump mechanism.
The logarithmic temperature dependence of the AH resistivity and
the absence of such  for the AH conductivity observed in amorphous
iron films~\cite{bergmannye} were initially attributed to the
Coulomb interactions~\cite{langenfeldwolfle} and  later
interpreted~\cite{dugaevea} in terms of the WL corrections for the
side-jump mechanism.
%

In a recent paper~\cite{mitra}, the logarithmic temperature
dependence of the longitudinal
and AH
resistivities of the polycrystalline iron films at sufficiently
low temperatures was reported. For well-conducting samples, the
behavior could be well explained by the WL
theory~\cite{langenfeldwolfle,dugaevea}  of the AHE in
two-dimensional homogeneously disordered samples. For more
resistive samples, however, noticeable deviations from the
theoretical predictions were observed. The authors suggested that
these deviations could be attributed to the granular structure of
the samples.

%

Motivated by the experimental data of Ref.~\onlinecite{mitra}, in
the present paper we investigate AHE in a granular system of
ferromagnetic metallic nanoparticles
within a microscopic theory.
For that purpose, we extended the recently developed
theory~\cite{KEprl,KEprb} of the conventional Hall effect in
granular metals to describe AHE.
%
%

The paper is organized as follows. In Sec.~\ref{sec:model}, we
formulate the model for a granular system. In
Sec.~\ref{sec:classical}, the residual AH resistivity is
calculated, first using the classical approach, and then this
result is recovered from the diagrammatic approach. The breakdown
of the scaling relation between the AH and longitudinal
resistivities is discussed. In Sec.~\ref{sec:WL}, we calculate the
WL corrections to the AH resistivity and discuss the experiment of
Ref.~\onlinecite{mitra}. Concluding remarks are  presented in
Sec.~\ref{sec:conclusion}.

\section{Model \label{sec:model}}

\begin{figure}
\includegraphics[width=\linewidth]{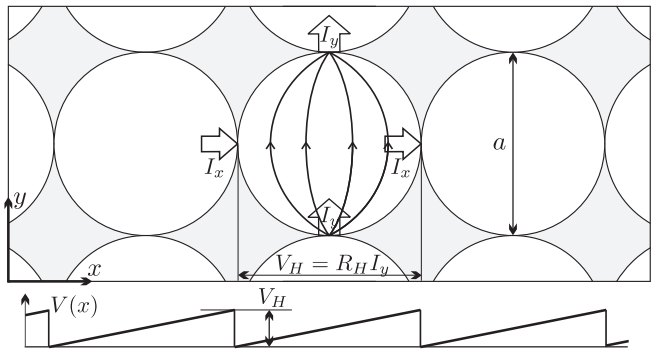}
\caption{\label{fig:system} Granular system and the classical
picture of the Hall conductivity. The external Ohmic voltage $V_y$
is applied to the contacts in the $y$ direction. The resulting
Ohmic current $I_y=G_T V_y$ running through the grain in the $y$
direction causes the Hall voltage drop $V_H=R_H I_y$ between its
opposite banks in the $x$ direction. Since when calculating the
Hall conductivity $\sig_{xy}$ the total voltage drop per lattice
period in the $x$ direction is assumed 0, the Hall voltage $V_H$
is applied with an opposite sign to the contacts in the $x$
direction (see bottom), causing the Hall current $I_x=G_T V_H =
G_T^2 R_H V_y$ [see Eq.~(\ref{eq:sigxy})].}
\end{figure}

We consider a regular quadratic ($d=2$, single granular monolayer)
or cubic ($d=3$, many monolayers) lattice of identical in form and
size three-dimensional metallic grains coupled to each other by
tunnel contacts with identical conductances $G_T$. At the same
time, we assume that the grains are disordered either due to
impurities in the bulk of the grains or due to an atomically
irregular shape.  The assumptions of the regularity of the system
simplify the analysis significantly, but are not crucial. The
results we obtain are expected to apply to structurally disordered
granular arrays as well.
We consider the metallic regime in this paper, when the tunnel
conductance $G_T=(2e^2/\hbar)g_T$ is much larger than the quantum
conductance,
    \beq
    \label{eq:gT}
        g_T\gg 1.
    \eeq
In this limit, the whole granular system is a good conductor and
the quantum effects of weak localization and Coulomb interactions
can be studied~\cite{BLVErmp} perturbatively in $1/g_T \ll 1$.
As usual, it is assumed that the granularity is
well-pronounced~\cite{BLVErmp}, i.e., the dimensionless grain
conductance $g_0$ exceeds the tunnel conductance $g_T$,
    \beq
            g_0 \gg g_T.
    \label{eq:g0gT}
    \eeq

The key ingredients of AHE that give rise to a  finite transversal
conductivity $\sig_{xy}$ are (i) the spin magnetization of
conduction electrons and (ii) considerable spin-orbit interaction.
Analogously to homogenously disordered metals, the simplest
Hamiltonian containing these two ingredients and thus describing
AHE in a granular system can be written as
    \beq
    \label{eq:H}
    \Hh = \Hh_0 + \Hh_U + \Hh_T.
    \eeq
The first two terms in Eq.~(\ref{eq:H}) describe  isolated grains,
where
\begin{subequations}
    \beq
    \label{eq:H0} \Hh_0 = \sum\limits_{\fati}\intr{\fatr_\fati}\,
  \htrps\Psi(\fatr_\fati)
  \left[-\frac{\nabla_{\fatr_\fati}^2}{2m}-\epsilon_F
     -\inner{\fath}{\fatsigma}
     \right]
  \Psi(\fatr_\fati)
    \eeq
contains the kinetic energy and the exchange field $\hb=h \eb_z$
directed along the $z$ axis in all the grains (we put $\hbar=1$).
The exchange field causes a finite spin magnetization of the
electrons. The intragrain disorder is described by
   \begin{eqnarray}
    \Hh_U &=& \sum\limits_{\fati}\intr{\fatr_\fati} \htrps\Psi(\fatr_\fati)
  \left\{
     U(\fatr_\fati) \right. \nonumber \\
     & & \left.
     - \ii \lambda^2\inner{\fatsigma}{\cross{\nabla U(\fatr_\fati)}{\nabla_{\fatr_\fati}}}
  \right\}
  \Psi(\fatr_\fati),
    \label{eq:HU}
    \end{eqnarray}
where the first term corresponds to the conventional scattering on
the disorder potential $U(\rb_\ib)$ and the second one to the
spin-orbit scattering. 
In Eqs.~(\ref{eq:H0}) and (\ref{eq:HU}), $\Psi =
\trps{\ztupel{\psi_\uparrow}{\psi_\downarrow}}$ is the
two-component spinor field operator of the electrons, $\fatsigma =
(\sigma_x,\sigma_y,\sigma_z)$ denotes the vector consisting of the
Pauli matrices $\sig_\al$, $\al=x,y,z$, and $\fati=(i_1, \ldots,
i_d)$ is an integer tuple numerating the grains on the lattice.
The integration with respect to $\rb_\ib$ is performed over volume
of the grain $\fati$.

We consider the simplest model of disorder
    \beq
        U(\rb)=\sum_a u \de(\rb-\rb_a),
    \label{eq:U}
    \eeq
in which the point impurities are located at random positions
$\rb_a$ within the grains and are uniformly distributed with the
concentration $n_\text{i}$ over the volume of the grains. We
assume that spin-orbit coupling is weak in the sense $\lambda p_\F
\ll 1$, where $p_F$ is the Fermi momentum, and that the exchange
field $h \ll \epsilon_F$ is smaller than the Fermi energy
$\epsilon_F$ of electrons in the grains. The latter two
assumptions allow one to study  AHE perturbatively in $h$ and
spin-orbit coupling.

The last term in the Hamiltonian (\ref{eq:H}) describes tunneling
between the grains,
    \begin{eqnarray}
            \Hh_T &=& \sum\limits_{\average{\ib,\jb}}
        (X_{\ib,\jb}+X_{\jb,\ib}), \nonumber \\
         X_{\ib,\jb} &=&
  \intr{\s_\ib}\dt \s_\jb\,
  t(\fats_\fati,\fats_\jb)
  \htrps\Psi(\fats_\fati)\Psi(\fats_\fatj).
    \label{eq:HT}
    \end{eqnarray}
\end{subequations}
The summation in Eq.~(\ref{eq:HT}) is done over the neighboring
grains $\ib$ and $\jb$, so that each tunnel contact is counted
only once, and the integration with respect to  $\s_\ib$ and
$\s_\jb$ is performed over two surfaces of the contact between the
grains $\ib$ and $\jb$, one belonging to grain~$\fati$ and the
other to grain~$\fatj$. It is both physically reasonable and
convenient for calculations~\cite{KEprb} to treat the tunneling
amplitudes~$t(\fats_\fati,\fats_\fatj)$ as Gaussian random
variables with the
variance~$\taverage{t(\fats_\fati,\fats_\fatj)t(\fats_\fatj,\fats_\fati)}_t
= t_0^2\,\delta(\fats_\fati-\fats_\fatj)$.

The anomalous Hall conductivity of the array is calculated using
the Kubo formula for granular systems in the Matsubara
representation,
\beq
    \sig_{\ab\bb}(\om)=  a^{2-d}\frac{1}{\om}
        \lt[ \Pi_{\ab\bb}(\om)-\Pi_{\ab\bb}(0) \rt],
\label{eq:Kubo}
    \eeq
where
    \beq
    \Pi_{\ab\bb}(\om)= -\sum_\jb 
        \int_0^{1/T} \mathrm{d}\tau\,\mathrm{e}^{\ii \om \tau}
        \lan \mathrm{T}_\tau  [I_{\ib,\ab}(\tau) I_{\jb,\bb}(0)]\ran
    \label{eq:Pi}
    \eeq
is the correlation function of the
tunnel currents
    \beq
        I_{\ib,\ab} =-\ii e (X_{\ib+\ab,\ib}-X_{\ib,\ib+\ab}).
    \label{eq:I}
    \eeq
Here, $\om $ is a bosonic Matsubara frequency (we assume $\om>0$
throughout the paper), the lattice unit vectors $\ab$ and $\bb$
denote the directions of the current and  external electric field,
respectively. The approach to calculating the AH conductivity is
analogous to that developed for the ordinary Hall effect in
Ref.~\onlinecite{KEprb}. It is based on the diagrammatic
perturbation theory in the tunnel Hamiltonian (\ref{eq:HT}) with
the ratio $g_T/g_0$ [Eq.~(\ref{eq:g0gT})] of the tunnel and grain
conductances as an expansion parameter. We refer the reader to
Ref.~\onlinecite{KEprb} for the details of the approach.

In our model (\ref{eq:H})-(\ref{eq:HT}), the source of spin-orbit
scattering are the impurities in the bulk of the grains [second
term in Eq.~(\ref{eq:HU})].
The anomalous Hall current, therefore, arises from the bulk of the
grains.
To perform explicit calculations, we will assume the intragrain
dynamics is diffusive, i.e., the mean free path $l\ll a$ is much
smaller than the grain size $a$.
In the opposite case of clean grains, surface scattering is
dominant and spin-orbit scattering off the grain boundary could be
the major source of AHE.
If the boundary roughness can effectively be modeled by scattering
on impurities in the bulk of the grains, our results may also be
applicable to the arrays of ballistic grains with chaotic
intragrain dynamics.

\section{Residual anomalous Hall resistivity\label{sec:classical}}
\subsection{Classical approach}


We start by calculating the residual anomalous Hall conductivity
$\sig_{xy}$ and resistivity $\rho_{xy}$ of a granular array,
neglecting quantum effects of weak localization and Coulomb
interactions.
Actually, as we show in this subsection, as long as quantum
effects are neglected the AH conductivity can be obtained by means
of the classical electrodynamics without using the Kubo formula.
The  diagrammatic approach that will further allow us to include
quantum effects is presented in Sec.~\ref{sec:classicaldiagram}.

Within the classical approach, the granular array can be
considered as a resistor network with the tunnel contacts  viewed
as surface resistors with conductance $G_T$. The AHE occurs inside
the grains and is fully characterized by the AH resistance $R_H$
of the each grain. Given $R_H$, in the leading order in $g_T / g_0
\ll 1$, one can easily arrive (Fig.~\ref{fig:system}) at the
expression
    \beq
        \sig_{xy}= a^{2-d}G_T^2 R_H
        \label{eq:sigxy}
    \eeq
for the residual AH conductivity of the granular array. Since the
longitudinal conductivity equals
    \beq
        \sig_{xx}=a^{2-d} G_T,
    \label{eq:sigxx}
    \eeq
for the AH resistivity of the granular system we obtain
    \beq
            \rho_{xy}=\frac{\sig_{xy}}{\sig_{xx}^2}= a^{d-2} R_H.
    \label{eq:rhoxy}
    \eeq
The AH resistivity is, therefore, expressed solely through the
Hall resistance $R_H$ of a single grain and is independent of the
tunnel conductance $G_T$, which determines the longitudinal
resistivity [Eq.~(\ref{eq:sigxx})].

To get  a further insight into the problem, one should specify
$R_H$ more explicitly. The electron transport in the diffusive
grains can fully be described by the specific longitudinal
$\sig_{xx}^\text{gr}$ and AH $\sig_{xy}^\text{gr}$ conductivities
of the grain material. The AH conductivity
    \beq
        \sig_{xy}^\text{gr}=\sig_{xy}^\text{gr,ss}+\sig_{xy}^\text{gr,sj}
    \label{eq:sigxygr}
    \eeq
is a sum of two contributions due to skew-scattering (ss) and
side-jump (sj) mechanisms. Given $\sig_{xx}^\text{gr}$ and
$\sig_{xy}^\text{gr}$, one can find the anomalous Hall resistance
$R_H$ of the grain by solving the  electrodynamics problem for the
distribution of the electric potential in the grain~\cite{KEprb}.
%
%
%
%
Analyzing this problem, one obtains that $R_H$ is expressed
through the specific AH resistivity of the grain material
$\rho_{xy}^\text{gr}=\sigma_{xy}^\text{gr}/(\sigma_{xx}^\text{gr})^2$
in the following way
    \beq
    \label{eq:RH}
        R_H = A_H \rho_{xy}^\text{gr}/a,
    \eeq
where the numerical factor $A_H \leq 1$ is determined by the shape
of the grain only. For simple grain geometries $A_H$ can be found
explicitly, e.g., $A_H=1$ for cubic and $A_H=\pi/4$ for spherical
grains.

As follows from Eqs.~(\ref{eq:rhoxy}) and (\ref{eq:RH}), the AH
resistivity of a three-dimensional granular array ($d=3$, many
grain monolayers)
    \beq
            \rho_{xy}=A_H \rho_{xy}^\text{gr}
    \label{eq:rhoxy3D}
    \eeq
is determined by the AH resistivity of the grain material, up to a
geometrical numerical factor determined by the shape of the grain.
The AH resistivity of a granular film ($d=2$, one to several
monolayers)
    \beq
            \rho_{xy}=A_H \rho_{xy}^\text{gr} / d_z
    \label{eq:rhoxy2D}
    \eeq
is obtained by dividing the 3D result (\ref{eq:rhoxy3D}) by the
thickness $d_z$ of the film.

The results (\ref{eq:sigxy})-(\ref{eq:rhoxy2D}) are actually
analogous to those obtained in Refs.~\onlinecite{KEprl} and
\onlinecite{KEprb} for the conventional Hall effect in granular
metals, with the AH resistivity of the grain material
$\rho_{xy}^\text{gr}$ entering the equations instead of the
conventional Hall resistivity. Specifics of AHE is reflected in,
e.g., the breakdown of the scaling relation between the AH and
longitudinal resistivities,
as discussed in the next subsection.

\subsection{Breakdown of the scaling relation}


In homogeneously disordered systems, for certain types of disorder
the AH and longitudinal resistivities obey the scaling relation
    \beq
        \rho_{xy}\propto\rho_{xx}^\gamma
    \label{eq:scaling}
    \eeq
with the exponent $\ga=1$  for skew-scattering and $\ga=2$ for
side-jump mechanisms. This scaling originates from the fact that
spin-orbit scattering, which results in the transversal current,
is caused by  the same impurity potential $U(\rb)$
[Eq.~(\ref{eq:HU})], scattering off which is responsible for the
finite longitudinal resistivity.

The scaling relation (\ref{eq:scaling}) holds for the model of
identical randomly placed short-range impurities, described by
Eqs.~(\ref{eq:HU}) and (\ref{eq:U}). Within this model,
the longitudinal
and AH conductivities of the grain material equal
    \beq
            \sig_{xx}^\text{gr}=
            \frac{e^2 v_F^2}{3 \pi n_\text{i} u^2},
    \label{eq:sigxxgr}
    \eeq
    \beq
            \sig_{xy}^\text{gr,ss}=
            -\frac{5\pi}{3} \nu u (\la p_F)^2
        \frac{h}{\epsilon_F} \sig_{xx}^\text{gr},
    \label{eq:sigxygrss}
    \eeq
    \beq
            \sig_{xy}^\text{gr,sj}=-3\pi (\nu u)^2 \frac{n_\text{i}}{\nu \epsilon_F}(\la p_F)^2
            \frac{h}{\epsilon_F} \sig_{xx}^\text{gr}.
    \label{eq:sigxygrsj}
    \eeq
Here, $\nu$ is the
density of states at the Fermi level for $h=0$.
%
As seen from Eqs.~(\ref{eq:sigxxgr})-(\ref{eq:sigxygrsj}), as  the
impurity concentration $n_\text{i}$ is varied, the resistivities
$\rho_{xx}^\text{gr}=1/\sig_{xx}^\text{gr}$ and
$\rho_{xy}^\text{gr}$ indeed change according to
Eq.~(\ref{eq:scaling}) [it is  implied in Eq.~(\ref{eq:scaling})
that the variation of $\rho_{xy}^\text{gr}$ and
$\rho_{xx}^\text{gr}$ is caused by the change of $n_\text{i}$,
i.e., the amount of disorder, whereas the strength $u$ of the
scattering potential of single impurities is fixed]. The scaling
relation (\ref{eq:scaling}) thus holds for the specific
resistivities of the grain material. Although
Eqs.~(\ref{eq:sigxxgr})-(\ref{eq:sigxygrsj}) are obtained for weak
impurity scattering (Born approximation), it can be
shown~\cite{MW} that the scaling law (\ref{eq:scaling}) holds for
strong scattering as well, since the dependence on the impurity
concentration remains the same. However, for more complicated type
of disorder with stronger finite-range correlations of the
disorder potential
the scaling relation may be violated.

Comparing Eqs.~(\ref{eq:sigxx}) and (\ref{eq:rhoxy}), we see that
no scaling relation similar to (\ref{eq:scaling}) between the AH
$\rho_{xy}$ and longitudinal
    \beq
        \rho_{xx}=a^{d-2}/G_T
    \label{eq:rhoxx}
    \eeq
resistivities of the whole granular array holds. This result is
actually not surprising, since the longitudinal and AH transport
in granular systems are governed by different mechanisms: the
former is due to tunneling through the potential barriers between
the grains, whereas the latter is caused by spin-orbit scattering
inside the grains.
%
%
%
According to Eqs.~(\ref{eq:sigxx}) and (\ref{eq:rhoxy}), if the
granularity is indeed pronounced [Eq.~(\ref{eq:g0gT})], the AH
resistivity should not vary much for samples with noticeably
different longitudinal resistivities. One could say that for
granular systems, the scaling relation (\ref{eq:scaling}) with the
exponent $\ga=0$ holds, independent of the dominant mechanism of
AHE.
%
%
In this context, we note that considerable deviations from the
scaling law (\ref{eq:scaling}) have previously been observed
experimentally in several types of heterostructure systems
\cite{xiongetal,xuetal}, in which the longitudinal resistivity was
also governed by the structural disorder (such as transparency of
the interfaces) rather than by the intrinsic disorder of the
ferromagnetic material.


\subsection{Residual anomalous Hall resistivity via diagrammatic approach \label{sec:classicaldiagram}}

The classical approach allows one to easily obtain
Eq.~(\ref{eq:sigxy}) for the residual AH conductivity and make
some interesting conclusions about AH transport in granular metals
at high enough temperatures. However, it has
nothing to say about quantum effects of weak localization and
Coulomb interactions, which set in at sufficiently low
temperatures. To study these effects on the AH transport, a more
sophisticated diagrammatic approach based on the Kubo formula
(\ref{eq:Kubo}) is needed.
Before we proceed with the quantum effects in Sec.~\ref{sec:WL},
we first demonstrate here how the classical result
(\ref{eq:sigxy}) is reproduced within the diagrammatic approach.

As demonstrated in Ref.~\onlinecite{KEprb}, the key object of the
diagrammatic approach to the Hall effect in granular systems is
the {\em intragrain diffuson}, i.e., the two-particle electron
propagator of an isolated grain. It contains all the information
about the specific mechanism of the Hall effect.
As usual, the  diffuson is formally defined as the
disorder-averaged product of two Greens's functions. In the
presence of the exchange field and spin-orbit scattering the
electron Green's functions are matrices in the spin space and the
intragrain diffuson is defined as their direct product,
    \beq
     \hat{D}_\om(\rb,\rb') =
 \frac{1}{2\pi\nu}
 \lan \hat{\Gc}_{\eps+\om}(\rb,\rb') \otimes
           \hat{\Gr}_\eps(\rb',\rb) \ran_U, \mbox{ }(\eps+\om)\eps < 0
    \eeq
Here, $\hat{\Gc}$'s are the exact Green's functions
of the intragrain Hamiltonian $\Hh_0+\Hh_U$ in the Matsubara
technique  for a given realization of the disorder potential
$U(\rb)$ and the angle brackets $\lan \ldots \ran_U$ denote
disorder-averaging.

According to the Kubo formula (\ref{eq:Kubo}), the conductivity is
in the leading order expressed through the spin-singlet diffuson
component, which is given by the trace~$\text{Tr}_\sig$  of the
Green's functions in the spin space
    \beq D_\omega(\fatr,\fatr') =
 \frac{1}{4\pi\nu}
 \lan \text{Tr}_\sig [\hat{\Gc}_{\eps+\om}(\rb,\rb')
           \hat{\Gc}_\eps(\rb',\rb)]\ran_U, \mbox{ }(\eps+\om)\eps<
           0.
    \label{eq:Ddef}
    \eeq
 Below, we will need the spin-singlet diffuson (\ref{eq:Ddef}) only.

\begin{figure}
\includegraphics[width = 0.75\linewidth]{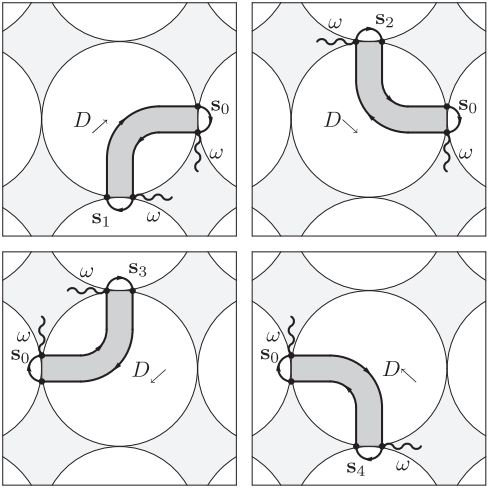}
\caption{Diagrams for the residual anomalous Hall conductivity
$\sig_{xy}$ [Eqs.~(\ref{eq:sigxy}) and (\ref{eq:RHexpr})], see
Ref.~\onlinecite{KEprb} for details.} \label{fig:sigxy}
\end{figure}

Analogously to the conventional Hall effect~\cite{KEprl,KEprb},
the residual AH conductivity $\sig_{xy}$ is given by the diagrams
in Fig.~\ref{fig:sigxy}. Calculating these diagrams, one can
relate the tunnel conductance in Eq.~(\ref{eq:sigxy}) for
$\sig_{xy}$ to the microscopic parameters of the model as
$g_T=2\pi (\nu t_0)^2 S_0$ ($S_0$ is the area of the contact) and
express the AH resistance of the grain through to the intragrain
diffuson (\ref{eq:Ddef}) as
    \beq
        R_H=\frac{1}{2e^2 \nu} (D_\nea -D_\sea + D_\swa - D_\nwa),
    \label{eq:RHexpr}
    \eeq
where
    \beq
    D_\al=\frac{1}{S_0^2} \int \dt \s_0 \dt\s_a D_{\om=0}(\s_0,\s_a)
    \label{eq:Daldef}
    \eeq
are the diffusons at zero frequency $\om=0$ connecting different
contacts as shown in Fig.~\ref{fig:sigxy}, with $a=1,2,3,4$ for
$\al=\nea,\sea,\swa,\nwa$, respectively.

The problem of calculating $R_H$ is, therefore, reduced to finding
the diffuson. Within the conventional disorder-averaging
technique~\cite{abrikosov}, the
diffuson (\ref{eq:Ddef}) can be shown to satisfy the diffusion
equation
    \beq \label{eq:Deq}
        (\omega - D_0\nabla_\fatr^2) D_\omega(\fatr,\fatr') = \delta(\fatr-\fatr'),
    \eeq
in which $D_0= v_F^2 \tau /3 $ is the coefficient of the
intragrain diffusion ($v_F$ is the Fermi velocity and $\tau$ is
the scattering time, $1/\tau=2\pi \nu n_\text{i} u^2$).
%
Equation (\ref{eq:Deq}) itself clearly does not contain any
information about the Hall effect. This information is contained
in the boundary condition for $D_\om(\rb,\rb')$, which
Eq.~(\ref{eq:Deq}) must be supplied with for a finite system.
In Ref.~\onlinecite{KEprb} a general method of deriving the
boundary condition for the diffuson was developed and
it was shown that the  boundary condition may be written
as
    \beq
        n_\al \lan j_\al r_\be \ran \nabla_{\rb\be}
        D_\om(\rb,\rb')|_{\rb\in S} = 0.
    \label{eq:Dbc}
    \eeq
Here, $\al,\be=x,y,z$, and $n_\al$ are the components of the unit
vector $\fatn$ normal to the grain boundary $S$ at point $\rb$ and
pointing out of the grain. In Eq.~(\ref{eq:Dbc}),
    \beq
            \lan j_\al r_\be \ran
        = \frac{1}{2}\text{Tr}_\sig \int \dt\rb'\,  \lan \hat{j}_{\rb\al} [\Gc_{\eps+\om} (\rb,\rb')
        \Gc_\eps(\rb',\rb)] (\rb'-\rb)_\be \ran_U
    \label{eq:jr}
    \eeq
is the current-coordinate correlation function. The
nonrelativistic part of the current operator $\hat{\jb}$ has the
conventional form
    \beq
            \hat{\jb}[\psi^*(\rb) ,\psi(\rb)]=\frac{-\ii}{2m}[\psi^*(\rb) \nabla \psi(\rb)-
            \psi(\rb) \nabla \psi^*(\rb)]
    \label{eq:j}
    \eeq

Explicit form of the boundary condition (\ref{eq:Dbc}) is thus
determined solely by $\lan j_\al r_\be \ran$. Analogously to the
conductivity tensor, only the longitudinal $\lan j_x r_x \ran=
\lan j_y r_y \ran=\lan j_z r_z \ran$ and Hall $\lan j_x r_y \ran=
-\lan j_y r_x \ran$ components are nonzero (we remind the reader
that the exchange field is directed along the $z$ axis). This
allows us to rewrite Eq.~(\ref{eq:Dbc}) in the form
    \beq
            \lan j_x r_x \ran (\nb \cdot \nabla_{\rb} D_\om(\rb,\rb'))|_{\rb \in S}=\lan j_x r_y \ran
            (\tb \cdot \nabla_{\rb} D_\om(\rb,\rb'))|_{\rb \in S},
    \label{eq:Dbc2}
    \eeq
 where the vector $\tb =[\nb \times
\hb]/h$ is tangent the grain boundary at point $\rb$.

Since the AHE is weak due to the smallness of the spin-orbit
coupling constant $\la p_F \ll 1$ and the exchange field
$h/\epsilon_F\ll 1$, the longitudinal component $\lan j_x r_x
\ran$ can be calculated neglecting the exchange field and
spin-orbit scattering completely and the expression for it reads
    \beq
        \average{j_x r_x}=-\frac{4\pi}{3}\nu l^2.
    \label{eq:jxrx}
    \eeq

\begin{figure}
\includegraphics[width = 0.75\linewidth]{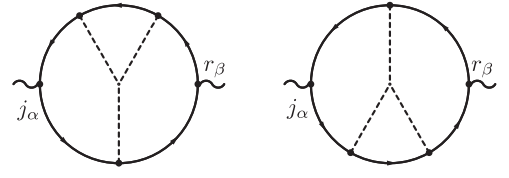}
\caption{\label{fig:jrss} Diagrams for the Hall component $\lan
j_x r_y \ran^\text{ss}$ of the current-coordinate correlation
function (\ref{eq:jr}) due to the skew-scattering mechanism.}
\end{figure}

All specifics of the AHE is contained in the Hall component $\lan
j_x r_y \ran$. Analogously to the AH conductivity of a
homogeneously disordered metal [see Eqs.~(\ref{eq:sigxygr}),
(\ref{eq:sigxygrss}) and (\ref{eq:sigxygrsj})], the total Hall
correlation function
    \beq
        \lan j_x r_y \ran = \lan j_x r_y \ran^\text{ss} + \lan j_x r_y
        \ran^\text{sj}
    \label{eq:jxry}
    \eeq
is the sum of two contributions due to skew-scattering (ss) and
side-jump (sj) mechanisms.

The skew-scattering part~$\average{j_xr_y}^\mathrm{ss}$ is given
by the diagrams in Fig.~\ref{fig:jrss}, which contain the impurity
lines describing the third-order scattering processes on a single
impurity, see, e.g., Ref.~\onlinecite{wolflemuttalib}. Calculating
these diagrams, we obtain
    \beq
        \label{eq:jrss}
        \average{j_x r_y}^{\mathrm{ss}}= -\frac{5\pi}{3} \nu u (\la
        p_F)^2
        \frac{h}{\epsilon_F}\average{j_x r_x}.
    \eeq


\begin{figure}
\includegraphics[width = 0.75\linewidth]{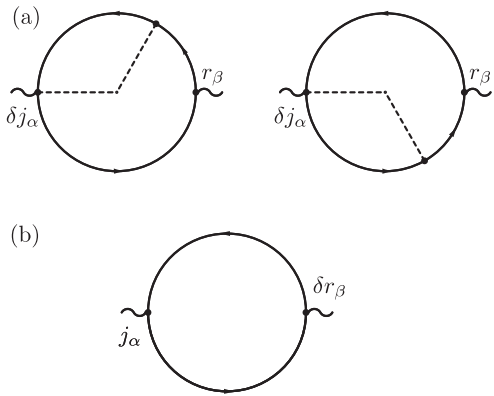}
\caption{\label{fig:jrsj} Diagrams for Hall component $\lan j_x
r_y \ran^\text{sj}$ of the current-coordinate correlation function
(\ref{eq:jr}) due to the side-jump mechanism. Diagrams~(a) and (b)
contain the relativistic corrections to the current [$\de
\hat{\jb}$, Eq.~(\ref{eq:dj})] and coordinate [$\de \hat{\rb}$,
Eq.~(\ref{eq:dr})] vertices, respectively.}
\end{figure}


The diagrams for the side-jump contribution $\average{j_x
r_y}^\mathrm{sj}$ are shown in Fig.~\ref{fig:jrsj}. The diagram in
Fig.~\ref{fig:jrsj}~(a) contains the conventional for the
side-jump mechanism relativistic correction
    \beq
        \de \hat{\jb} =
        \lambda^2\cross{\fatsigma}{\nabla
        U(\fatr)},
    \label{eq:dj}
    \eeq
to the current operator (\ref{eq:j}), see, e.g.,
Ref.~\onlinecite{wolflemuttalib}. Additionally, there exists an
analogous relativistic correction to the coordinate vertex. This
contribution can be obtained by repeating the derivation of the
boundary condition (\ref{eq:Dbc}) done in Ref.~\onlinecite{KEprb},
but taking into account the spin-orbit term of $\Hh_U$
[Eq.~(\ref{eq:HU})] in the diffuson ladder. This gives the diagram
in Fig.~\ref{fig:jrsj}~(b), in which
    \beq
    \label{eq:dr} \de \hat{\rb} =
    \la^2 [\fatsigma \times \hat{\pb} ]
    \eeq
is the relativistic correction to the coordinate operator
($\hat{\pb}=-\text{i} \nabla$). One can recognize that $\de
\hat{\rb}$ is the operator of the lateral translation
(``side-jump''), see, e.g. Ref.~\onlinecite{crepieuxbruno}.
Calculating the diagrams in Fig.~\ref{fig:jrsj},
we obtain
    \beq
    \label{eq:jrsj}
        \average{j_x r_y}^\mathrm{sj}
        =-3\pi (\nu u)^2 \frac{n_\text{i}}{\nu \epsilon_F}(\la p_F)^2 \frac{h}{\epsilon_F} \average{j_x r_x}.
    \eeq

Comparing Eqs.~(\ref{eq:jxrx}), (\ref{eq:jrss}) and
(\ref{eq:jrsj}) with Eqs.~(\ref{eq:sigxxgr}), (\ref{eq:sigxygrss})
and (\ref{eq:sigxygrsj}), we note that for both skew-scattering
and side-jump contributions the relation
    \beq
        \frac{\average{j_x r_y}^\mathrm{ss/sj}}{\average{j_x r_x}} =
        \frac{\sig^{\gr,\mathrm{ss/sj}}_{xy}}{\sig^\gr_{xx}}
    \eeq
holds.
%
Therefore the boundary condition (\ref{eq:Dbc}) may be rewritten
as
    \beq
            \sig_{xx}^\text{gr} (\nb \cdot \nabla_{\rb}D_\om(\rb,\rb'))|_{\rb \in S}=\sig_{xy}^\text{gr}
            (\tb \cdot \nabla_{\rb}D_\om(\rb,\rb'))|_{\rb \in S}.
    \label{eq:Dbc3}
    \eeq
As shown in Ref.~\onlinecite{KEprb}, it is exactly this form of
the boundary condition, which is necessary to reproduce the
classical result (\ref{eq:RH}) for the Hall resistance $R_H$
obtained by solving the electrodynamics problem.

Having established the correspondence between the classical and
diagrammatic approaches, comparing Eqs.~(\ref{eq:RHexpr}),
(\ref{eq:Deq}), and (\ref{eq:Dbc2}) with Eq.~(\ref{eq:RH}) and
using  the Einstein relation $\sig_{xx}^\text{gr}=2 e^2 \nu D_0$,
we can express the Hall resistance of the grain as
    \beq
            R_H = A_H \frac{1}{2 e^2 \nu} \frac{\lan j_x r_y \ran}{\lan j_x r_x
            \ran} \frac{1}{D_0 a}
    \label{eq:RHexpr2}
    \eeq
This form will be used in the next section for calculating WL
corrections.

\section{Weak localization corrections \label{sec:WL}}
\subsection{Calculations}

\begin{figure*}
\includegraphics[width=0.7\textwidth]{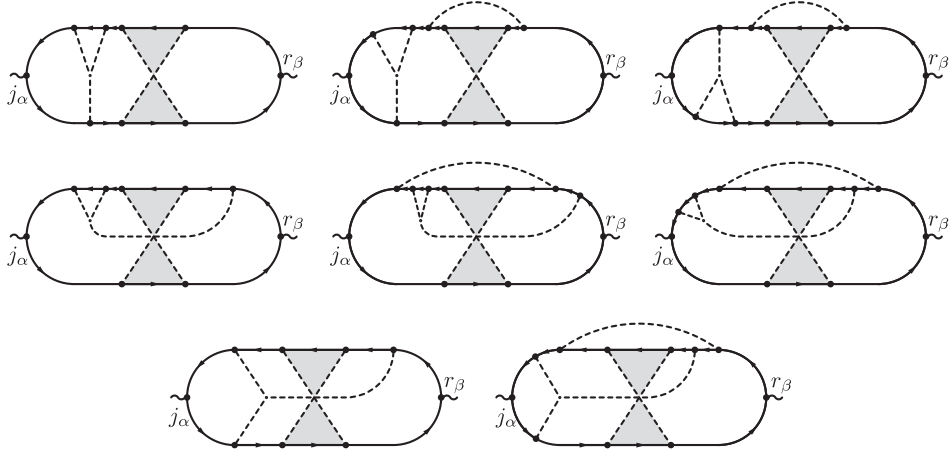}
\caption{\label{fig:djrss} Diagrams for the weak localization
correction $\de \lan j_x r_y \ran^\text{ss}$ to the
skew-scattering Hall component
$\taverage{j_xr_y}^\mathrm{ss}$~[Eq.~(\ref{eq:jrss})] of the
current-coordinate correlation function (\ref{eq:jr}). The gray
regions denote the Cooperons~[Eq.~(\ref{eq:C})]. Each diagram
depicted above stands for four diagrams: the ones not shown are
obtained by reflecting the impurity lines through the diagram
center and/or flipping the diagram upside-down.}
\end{figure*}
\begin{figure*}
\includegraphics[width=0.7\textwidth]{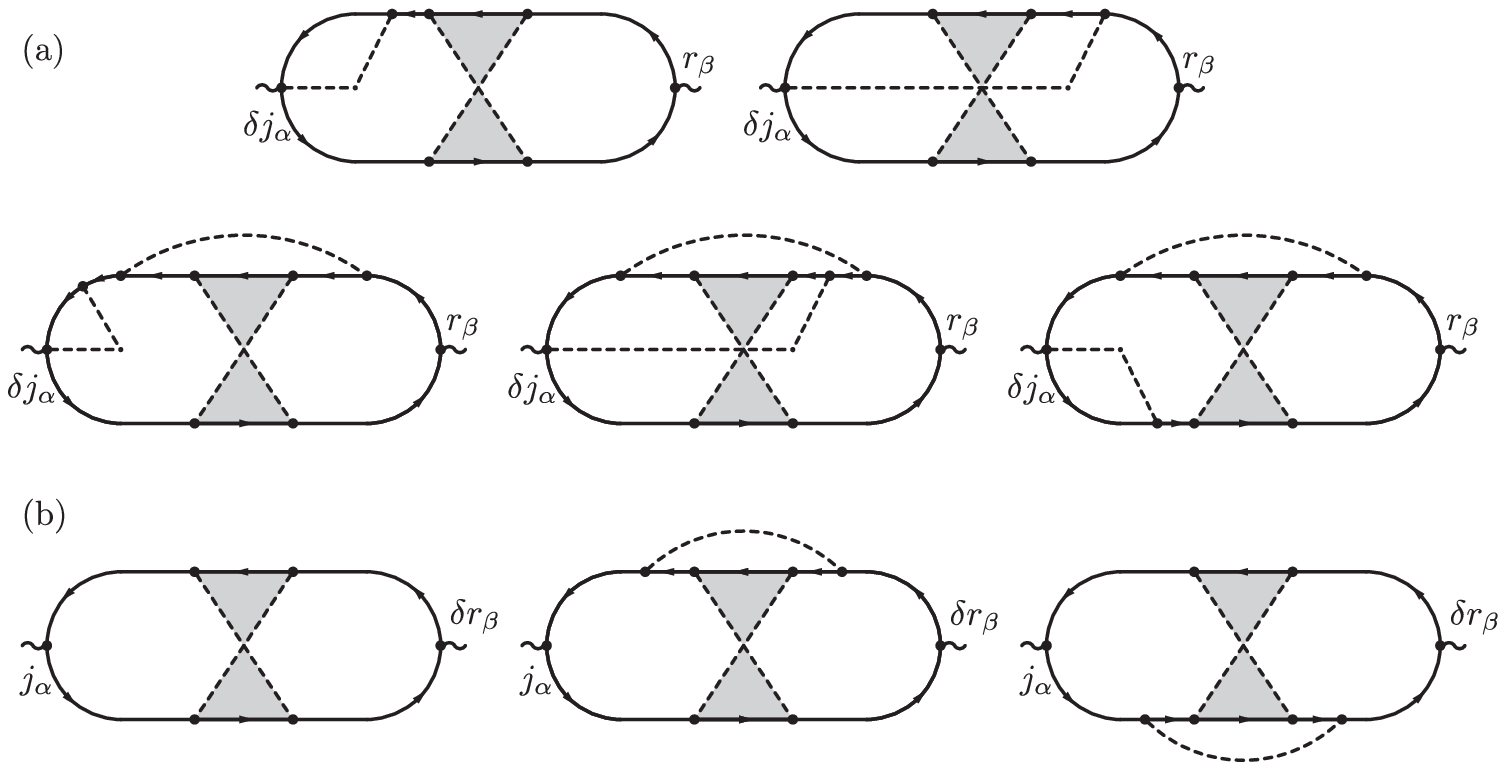}
\caption{\label{fig:djrsj} Diagrams for the weak localization
correction $\de \lan j_x r_y \ran^\text{ss}$ to the side-jump Hall
component $\taverage{j_xr_y}^\mathrm{sj}$~[Eq.~(\ref{eq:jrsj})] of
the current-coordinate correlation function (\ref{eq:jr}).
Diagrams~(a) and (b) contain the relativistic corrections to the
current [$\de \hat{\jb}$, Eq.~(\ref{eq:dj})] and coordinate [$\de
\hat{\rb}$, Eq.~(\ref{eq:dr})] vertices, respectively. Each
diagram in panel (a) stands for two diagrams: the one not shown is
obtained by flipping the diagram upside-down.}
\end{figure*}

We now proceed with calculating the weak localization corrections
to the obtained ``classical'' anomalous Hall conductivity
(\ref{eq:sigxy}) and resistivity (\ref{eq:rhoxy}) of the granular
metal.

Technically, one has to consider WL corrections to the diagrams in
Fig.~\ref{fig:sigxy} for the bare Hall conductivity
(\ref{eq:sigxy}) by inserting the Cooperon ladders into them in
all possible ways. As shown in Ref.~\onlinecite{KEprb}, such WL
corrections are factorized according to the form
Eq.~(\ref{eq:sigxy}), i.e., there are diagrams describing the
corrections to the tunnel conductance $G_T$ only and to the Hall
resistance $R_H$ of the grain only. This allows one to write down
the total weak localization correction $\de \sig_{xy}$ to AH
conductivity $\sig_{xy}$ in the form
    \beq
        \frac{\de \sig_{xy}}{\sig_{xy}}= 2 \frac{\de G_T}{G_T} +
        \frac{\de R_H}{R_H}
    \eeq
Naturally, the WL correction $\de G_T$ to the tunneling
conductance has the same form as that to the longitudinal
conductivity $\sig_{xx}$ [Eq.~(\ref{eq:sigxx})]~\cite{BLV,biagini}
and reads
    \beq
            \frac{\de G_T}{G_T} =\frac{\de \sig_{xx}}{\sig_{xx}} =
            \frac{\De}{2\pi}[C_{\om=0}(\ib+\ab,\ib)+C_{\om=0}(\ib,\ib+\ab)]
    \eeq
Here,
    \beq
        C_\om(\ib,\jb) =
            \int \frac{a^d \dt^d \fatq}{(2 \pi)^d}
            \frac{\mathrm{e}^{\ii a (\qb \cdot (\ib-\jb))} }{\omega + 2 \Ga \sum_{\alpha}[1-\cos(q_\al a)]+1/\tau_\varphi}
    \label{eq:C}
    \eeq
is the Cooperon of the whole granular array calculated in the
zero-mode approximation for the intragrain
Cooperons~\cite{BEAH,BLV,biagini}, $\De$ is the mean level spacing
in the grain, and $\ab=\eb_x$ or $\ab=\eb_y$. In Eq.~(\ref{eq:C}),
$\Ga=g_T \De$ is the tunneling rate, the dephasing time
$\tau_\varphi$ was  introduced by hand, and the integration with
respect to the quasimomentum $\qb$ is performed over the first
Brillouin zone $\qb \in [-\pi/a,\pi/a]^d$ of the grain lattice.
In order not to complicate the analysis, we assumed in
Eq.~(\ref{eq:C}) that the dephasing rate $1/\tau_\varphi \gg
1/\tau_\text{so}$ exceeds the spin orbit scattering rate
$1/\tau_\text{so}$. If $1/\tau_\varphi$, $1/\tau_\text{so}$, and
$h$ are of the same order, the spin structure of the Cooperon can
be taken into account as, e.g., in Ref.~\onlinecite{VG}.

According to Eq.~(\ref{eq:RHexpr2}), the AH resistance of the
grain has been expressed through the diffusion coefficient $D_0$
and the correlation functions $\lan j_x r_x \ran$ and $\lan j_x
r_y \ran$, which fully characterize the intragrain diffuson
$D_\om(\rb,\rb')$~[Eqs.~(\ref{eq:Deq}) and (\ref{eq:Dbc})]. As
these three are well-defined correlation functions,
one can calculate the WL corrections $\de D_0$, $\de \lan j_x r_x
\ran$, and $\de\lan j_x r_y \ran$ to them using the diagrammatic
technique.
%
This will allow us to obtain WL correction $\de \rho_{xy}$ to the
AH resistivity $\rho_{xy}$ from Eqs.~(\ref{eq:rhoxy}) and
(\ref{eq:RHexpr2}) as follows
    \beq
        \frac{\de \rho_{xy}}{\rho_{xy}}  =
        \frac{\de R_H}{R_H}= \frac{\de \lan j_x r_y \ran}{\lan j_x r_y \ran}-
        \frac{\de\lan j_x r_x \ran}{\lan j_x r_x \ran}-\frac{\de
        D_0}{D_0}.
    \label{eq:drhoxyexpr}
    \eeq

The WL corrections to the diffusion constant~$D_0$ and
longitudinal current-coordinate correlation function~$\average{j_x
r_x}$ are identical to those in Ref.~\onlinecite{KEprb} and have
the form
    \beq
    \label{Eq:WLD0JRlong}
    \frac{\de D_0}{D_0} = \frac{\de \lan j_x r_x\ran }{\lan j_x r_x \ran} = -c.
    \eeq
Here,
    \begin{eqnarray}
        c &=&
        \frac{\De}{\pi}C_{\om=0}(\ib,\ib) \nonumber \\
         &=&\frac{\De}{\pi} \int \frac{a^d \dt^d \fatq}{(2 \pi)^d}
           \frac{1}{2 \Ga \sum_{\alpha}[1-\cos(q_\al
           a)]+1/\tau_\varphi}.
    \label{eq:c}
    \end{eqnarray}



All specifics of AHE is contained in the Hall component $\lan j_x
r_y \ran$. The diagrams describing the WL corrections to $\lan j_x
r_y \ran^\text{ss}$~[Eq.~(\ref{eq:jrss})] and $\lan j_x r_y
\ran^\text{sj}$~[Eq.~(\ref{eq:jrsj})] are obtained from the
diagrams in Figs.~\ref{fig:jrss} and \ref{fig:jrsj} by inserting
the Cooperon ladder into them in all possible ways.

Let us first consider  WL correction $\de \lan j_x r_y
\ran^\text{ss}$ to  the skew-scattering  correlation function
$\lan j_x r_y \ran^\text{ss}$. The diagrams for $\de \lan j_x r_y
\ran^\text{ss}$ are shown in Fig.~\ref{fig:djrss}. In total, there
are 32 diagrams. After a tedious but straightforward calculation,
we obtain
    \beq \label{eq:djrss}
        \frac{\delta\average{j_xr_y}^\mathrm{ss}}
      {\average{j_xr_y}^\mathrm{ss}} = -c.
    \eeq

The diagrams for the WL correction $\de \lan j_x r_y
\ran^\text{sj}$ to the side-jump correlation function $\lan j_x
r_y \ran^\text{sj}$ are shown in Fig.~\ref{fig:djrsj}. The total
number of these diagrams is 13. Calculating them, we find that the
contributions from all these diagrams cancel each other
identically, which results in a vanishing correction
    \beq
    \label{eq:djrsj}
        \de\lan j_x r_y \ran^\text{sj}=0.
    \eeq
As seen from Eqs.~(\ref{eq:djrss}) and (\ref{eq:djrsj}), the
results for WL correction differ for skew-scattering and side-jump
mechanisms. Inserting Eqs.~(\ref{Eq:WLD0JRlong}),
(\ref{eq:djrss}), and (\ref{eq:djrsj}) into
Eq.~(\ref{eq:drhoxyexpr}), for the WL correction to the AH
resistivity we obtain
    \beq
        \frac{\de \rho_{xy}^\text{ss}}{\rho_{xy}^\text{ss}}=c,
    \label{eq:drhoxyss}
    \eeq
    \beq
        \frac{\de \rho_{xy}^\text{sj}}{\rho_{xy}^\text{sj}}=2 c
    \label{eq:drhoxysj}
    \eeq
for skew-scattering and side-jump mechanisms, respectively.
%
%
The total AH resistivity $\rho_{xy}= \rho_{xy}^\text{ss}+
\rho_{xy}^\text{sj}$ is the sum of the skew-scattering and
side-jump contributions and for the total WL correction $\de
\rho_{xy}= \de \rho_{xy}^\text{ss}+ \de \rho_{xy}^\text{sj}$ one
obtains from Eqs.~(\ref{eq:drhoxyss}) and (\ref{eq:drhoxysj}) that
   \beq
        \frac{\de \rho_{xy}}{\rho_{xy}}= A_{xy} c,
    \label{eq:drhoxy}
    \eeq
where
    \beq
        A_{xy}=\frac{  \rho_{xy}^\text{ss}+ 2 \rho_{xy}^\text{sj}} {\rho_{xy}^\text{ss}+ \rho_{xy}^\text{sj}}.
    \label{eq:Axy}
    \eeq
The factor (\ref{eq:Axy}) equals $A_{xy}=1$  and $A_{xy}=2$ for prevailing skew-scattering ($\rho_{xy}^\text{ss} \gg
\rho_{xy}^\text{sj} $)  and side-jump ($\rho_{xy}^\text{ss} \ll
\rho_{xy}^\text{sj} $) mechanisms, respectively, and belongs to the range $1<A_{xy}<2$,  when two mechanisms give comparable contributions.

Equations (\ref{eq:drhoxy}) and (\ref{eq:Axy}), together
with Eq.~(\ref{eq:c}), constitute our final result for the weak
localization corrections to the anomalous Hall resistivity of a
granular metal. In the next subsection, we discuss the obtained
results.

\subsection{Discussion}

We note that the form of Eqs.~(\ref{eq:c}), (\ref{eq:drhoxyss}),
and (\ref{eq:drhoxysj}) agrees with the results for WL corrections
to the AH resistivity of homogeneous
metals~\cite{langenfeldwolfle,dugaevea} for both skew-scattering
and side-jump mechanisms. This is most clearly seen, when the main
contribution to the integral over $\qb$ in Eq.~(\ref{eq:c}) comes
from small momenta, $q a \ll 1$, or equivalently, from spatial
scales much exceeding the grain size. This happens in two ($d=2$)
and one ($d=1$) dimensions, the latter case of granular ``wires''
is, however, irrelevant for the Hall effect. In three dimensions
($d=3$), the integral over $\qb$ in Eq.~(\ref{eq:c}) converges, if
one neglects dephasing,  and the relative correction
$c(\tau_\varphi) \sim 1/g_T$, therefore, depends only weakly on
$\tau_\varphi$ for finite dephasing.

For two-dimensional  arrays (one to several grain monolayers),
neglecting dephasing, the integral with respect to $\qb$ in
Eq.~(\ref{eq:c}) is logarithmically   divergent  at  small momenta
$qa  \ll 1$.
This divergency is cut by the finite dephasing rate
$1/\tau_\varphi$. At low enough temperatures, when $1/\tau_\varphi
\ll \Ga$, the divergency is strong and one obtains
    \beq
    \label{eq:c2D}
        c =
        \frac{1}{2\pi g_\Box}\ln(\Ga \tau_\varphi),
    \eeq
where the dimensionless sheet conductance $g_\Box=\sig_{xx}/[2
e^2/(2 \pi \hbar)]=2\pi g_T$ of the array was introduced. At
higher temperatures, when the dephasing  rate $1/\tau_\varphi $
becomes of order or larger than the tunneling rate $\Ga$,
$1/\tau_\varphi \gtrsim \Ga$, the integral in Eq.~(\ref{eq:c}) is
not divergent at $q a \ll 1$ and the WL corrections of the
granular film are not logarithmic in $\tau_\varphi$ anymore. At
even higher temperatures, when the dephasing rate exceeds the
Thouless energy of the grain, $1/\tau_\varphi \gtrsim D_0/a^2$,
the contributions to WL corrections come from the bulk of each
single grain, whereas the coherence of the intergrain motion is
destroyed. Since in realistic granular systems the grains are
three-dimensional particles, the WL corrections in this case are
given by the result~\cite{gorkovlarkinkhmelnitskii} for a
three-dimensional sample,
\beq
       \frac{\de \rho_{xy}}{\rho_{xy}}=A_{xy} \frac{3\sqrt{3}}{8 (\epsilon_F
\tau)^2} \lt(\al-\sqrt{\frac{\tau}{\tau_\varphi}} \rt),
       \mbox{ } 1/\tau_\varphi \gg D_0/a^2.
\label{eq:drhoxy3D} \eeq Here, $\al\sim 1$ is a numerical
cutoff-dependent factor and $\tau$ is the intragrain scattering
time. The correction (\ref{eq:drhoxy3D}) has a conventional for
the 3D case square-root dependence on the dephasing rate. So, the
``large-scale'' low-temperature regime $1/\tau_\varphi \ll \Ga$ is
the only one, in which the WL corrections of a granular film are
logarithmic in $\tau_\varphi$.

Using Eq.~(\ref{eq:c2D}), we can write down the WL correction
(\ref{eq:drhoxy})
in the limit $1/\tau_\varphi \ll \Ga$
as
    \beq
        \frac{\de \rho_{xy}}{\rho_{xy}}=
        \frac{A_{xy}}{2\pi g_\Box}\ln(\Ga \tau_\varphi).
    \label{eq:drhoxyln}
    \eeq
In this form, the result (\ref{eq:drhoxyln}) is in full agreement
(up to a different infrared cutoff scale $\Ga$, which is
determined by the microscopic structure of the system) with that
for a conventional homogeneously disordered metal characterized by
the same sheet conductance $g_\Box$.
This sort of ``universality''
is actually quite expected, since WL corrections in 2D arise from
large mesoscopic spatial scales, at which the microscopic
structure of the material, whether it is homogeneous or granular,
becomes irrelevant.
Therefore, 
it would be impossible to distinguish between granular and
homogeneous two-dimensional material by measuring WL corrections.
In this context, we remind that  WL correction to the
longitudinal~\cite{BLV} and conventional Hall~\cite{KEprb}
resistivities of a granular metal have earlier been shown to agree
with those for homogeneously disordered metals.
In line with Eq.~(\ref{eq:drhoxyln}), one can write down WL
correction to the longitudinal resistivity of a granular film in
the form~\cite{BLV}
   \beq
        \frac{\de \rho_{xx}}{\rho_{xx}}=
        \frac{A_{xx}}{2 \pi g_\Box}\ln(\Ga \tau_\varphi).
    \label{eq:drhoxxln}
    \eeq
with $A_{xx}=1$.


In view of the obtained results, we would like to discuss the
recent experiment of Ref.~\onlinecite{mitra}. The authors of
Ref.~\onlinecite{mitra} reported on the logarithmic temperature
dependence of the longitudinal $\rho_{xx}(T)=R_{xx} \ln(T_0/T)$
and AH $\rho_{xy}(T)=R_{xy} \ln(T_0/T)$ resistivities of the
polycrystalline iron films at sufficiently low temperatures. For
the most conductive samples, the values of the prefactors $R_{xx}$
and $R_{xy}$
were in a good agreement with the theoretical
predictions~\cite{langenfeldwolfle,dugaevea} for WL corrections in
two-dimensional homogeneously disordered metals for the case of
the dominant skew-scattering (provided one assumes the linear
$1/\tau_\varphi(T) \propto T$ temperature dependence of the
dephasing rate, as predicted, e.g., for electron-electron
interactions by the diffusive Fermi liquid theory for both
homogeneous~\cite{altshuleraronovkhmelnitsky} and
granular~\cite{BLV} metals). This means that the $\ln
T$-dependencies of $\rho_{xx}(T)$ and $\rho_{xy}(T)$ were well
described by Eqs.~(\ref{eq:drhoxyln}) and (\ref{eq:drhoxxln}) with
$A_{xy} = 2$ (indicating that side-jump mechanism of AHE is
dominant in these samples) and $A_{xx}=1$.
This suggested the explanation of the observed behavior
in terms of  WL effects.
%
For more resistive samples the $\ln T$-dependence seemed to
persist, but the prefactors deviated significantly from the
predicted~\cite{langenfeldwolfle,dugaevea} values. That is, the
behavior of $\rho_{xx}(T)$ and $\rho_{xy}(T)$ could still be
described by Eqs.~(\ref{eq:drhoxyln}) and (\ref{eq:drhoxxln})
with, however, smaller prefactors $A_{xy} < 2$ and $A_{xx} < 1$.
%
%
%
The authors argued that these deviations could be explained by the
onset of granularity in  more resistive samples.
%
%

According to Ref.~\onlinecite{mitra}, in the regime of the
intragrain dephasing length $l_\varphi \ll a$ ($1/\tau_\varphi =
D_0/l_\varphi^2$) smaller than the grain size $a$, the WL
correction to the AH resistivity $\rho_{xy}$ had to be given by
Eq.~(\ref{eq:drhoxyln}), but with the grain conductance $g_0$
entering the denominator of the prefactor instead of the tunnel
conductance $g_T$,  $ \de \rho_{xy}/\rho_{xy} \propto
A_{xy}\ln(l_\varphi/l)/g_0$. This would indeed be the case for
flat pancake-shaped grains provided their 2D size $a$ were much
greater than their thickness $a_0$, so that $a_0 \ll l_\varphi \ll
a$. Considering that the most resistive samples in the experiment
of Ref.~\onlinecite{mitra} were about 2nm   thick, this would
require the grain size $a$ to be at least 20nm. However, the
authors of Ref.~\onlinecite{mitra} presented an estimate of the
tunneling rate $\Ga$ for 1nm grains,  which would correspond to
the case of 3D grains. In such a situation, the WL correction to
the AH resistivity should be described by Eq.~(\ref{eq:drhoxy3D})
and it is not logarithmic in $\tau_\varphi$. Moreover, using the
method developed in the present paper one can demonstrate that for
pancake grains in the regime $a_0 \ll l_\varphi \ll a$ the WL
correction to the longitudinal conductivity would be  a sum of the
logarithm and logarithm squared contributions in the dephasing
length,  $\de\sig_{xx} /\sig_{xx}\propto -g_T/g_0^2[
\ln^2(l_\varphi/l_0)+ \be \ln(l_\varphi/l)]$ ($\be \sim 1$ is a
geometrical factor, $l_0$ is the contact size).
This yields the $\ln^2 T$ -dependence of the longitudinal
resistivity $\rho_{xx}$, which does not seem to agree with the
data of Ref.~\onlinecite{mitra}, where both $\rho_{xx}$ and
$\rho_{xy}$ were logarithmic in temperature.
For these reasons, we do not think that the model of pancake
grains corresponds to the experimental situation of
Ref.~\onlinecite{mitra}.

At the same time, in the limit $1/\tau_\varphi \gg \Ga$, we have
demonstrated for the AH resistivity~[Eq.~(\ref{eq:drhoxyln})] and
it has been earlier shown~\cite{BLV} for the longitudinal
resistivity~[Eq.~(\ref{eq:drhoxxln})] that WL corrections are
essentially the same for granular and homogeneously disordered
metals. Since this is the only regime, in which the WL corrections
to both $\rho_{xx}$ and $\rho_{xy}$ of a granular film are
logarithmic in $\tau_\varphi$, we conclude that the observed
deviations of the pre\-factors $A_{xy}$ and $A_{xx}$ from the
values $A_{xy}=2$ and $A_{xx}=1$ cannot be explained by the
granular structure of the system and one should find an
alternative explanation of the effect.
We emphasize that in Ref.~\onlinecite{mitra} not only the
prefactor $A_{xy}$ for the Hall, but also for the
longitudinal resistivity deviated from its ``universal'' value
$A_{xx}=1$. Since AHE in the experiment of Ref.~\onlinecite{mitra} is
weak in the sense $\rho_{xy} \ll \rho_{xx}$,
the longitudinal resistivity $\rho_{xx}$ is not noticeably
affected by the Hall effect.
%
Therefore, the conclusion that the WL effects in a granular metal
cannot explain the observed behavior could be drawn based alone on
the earlier result (\ref{eq:drhoxxln}) for
the longitudinal resistivity, without any knowledge about AHE.

Let us also briefly discuss the role of the Coulomb interactions
in context of the data of Ref.~\onlinecite{mitra}. In
Refs.~\onlinecite{ET} and \onlinecite{BELV}, the Coulomb
interaction corrections to the longitudinal resistivity that are
specific to granular metals and absent in conventional metals were
found. Analogous corrections were shown to exist for the
conventional Hall resistivity in Ref.~\onlinecite{KEprb} and one
could demonstrate that the results of Ref.~\onlinecite{KEprb} also
apply to the AH resistivity. As these Coulomb interaction
corrections are logarithmic in temperature (in any dimensionality
of the array), one could be tempted to explain the data of
Ref.~\onlinecite{mitra} in terms of them. Unfortunately, this
would not be possible, since these corrections are of insulating
nature, i.e., the relative corrections to the resistivities
$\rho_{xx}$ and $\rho_{xy}$ are positive. Therefore, taking them
into account would increase the value of the prefactors in the
logarithmic $T$-dependencies of the AH and longitudinal
resistivities. This would be in contradiction with the data of
Ref.~\onlinecite{mitra}, where a decrease of the prefactors
$A_{xy}$ and $A_{xx}$ for more resistive samples was observed.

\section{Conclusion \label{sec:conclusion}}

In conclusion, we have theoretically investigated the anomalous
Hall effect in ferromagnetic granular metals. We found that
no scaling law relation between the residual anomalous Hall and
longitudinal resistivities of a granular metal holds, regardless
of whether this scaling holds  for the specific resistivities of
the grain material or not: the Hall resistivity of the whole array
does not change as the longitudinal resistivity of the array is
varied.
At the same time, the weak localization corrections to the
anomalous Hall resistivity of  two-dimensional granular metals are
found to be in full agreement with those for conventional metals.
This is explained by the fact that the weak localization effects
in low-dimensional conductors are determined by large mesoscopic
spatial scales, at which the microscopic structure of the system
is indistinguishable.


Financial support of SFB Transregio 12 is greatly appreciated.


\begin{thebibliography}{30}
\expandafter\ifx\csname
natexlab\endcsname\relax\def\natexlab#1{#1}\fi
\expandafter\ifx\csname bibnamefont\endcsname\relax
  \def\bibnamefont#1{#1}\fi
\expandafter\ifx\csname bibfnamefont\endcsname\relax
  \def\bibfnamefont#1{#1}\fi
\expandafter\ifx\csname citenamefont\endcsname\relax
  \def\citenamefont#1{#1}\fi
\expandafter\ifx\csname url\endcsname\relax
  \def\url#1{\texttt{#1}}\fi
\expandafter\ifx\csname
urlprefix\endcsname\relax\def\urlprefix{URL }\fi
\providecommand{\bibinfo}[2]{#2}
\providecommand{\eprint}[2][]{\url{#2}}

\bibitem[{\citenamefont{Karplus and Luttinger}(1954)}]{karplusluttinger}
\bibinfo{author}{\bibfnamefont{R.}~\bibnamefont{Karplus}} \bibnamefont{and}
  \bibinfo{author}{\bibfnamefont{J.~M.} \bibnamefont{Luttinger}},
  \bibinfo{journal}{Phys. Rev.} \textbf{\bibinfo{volume}{95}},
  \bibinfo{eid}{1154} (\bibinfo{year}{1954}).

\bibitem[{\citenamefont{W\"olfle and Muttalib}(2006)}]{wolflemuttalib}
\bibinfo{author}{\bibfnamefont{P.}~\bibnamefont{W\"olfle}} \bibnamefont{and}
  \bibinfo{author}{\bibfnamefont{K.}~\bibnamefont{Muttalib}},
  \bibinfo{journal}{Ann. Phys.} \textbf{\bibinfo{volume}{15}},
  \bibinfo{eid}{508} (\bibinfo{year}{2006}).

\bibitem[{\citenamefont{Sinitsyn}(2008)}]{Sinitsyn}
\bibinfo{author}{\bibfnamefont{N.~A.} \bibnamefont{Sinitsyn}},
  \bibinfo{journal}{J. Phys.: Condens. Matter} \textbf{\bibinfo{volume}{20}},
  \bibinfo{pages}{023201} (\bibinfo{year}{2008}).

\bibitem[{\citenamefont{Sundaram and Niu}(1999)}]{SN}
\bibinfo{author}{\bibfnamefont{G.}~\bibnamefont{Sundaram}} \bibnamefont{and}
  \bibinfo{author}{\bibfnamefont{Q.}~\bibnamefont{Niu}},
  \bibinfo{journal}{Phys. Rev. B} \textbf{\bibinfo{volume}{59}},
  \bibinfo{pages}{14915} (\bibinfo{year}{1999}).

\bibitem[{\citenamefont{Taguchi et~al.}(2001)\citenamefont{Taguchi, Oohara,
  Yoshizawa, Nagaosa, and Tokura}}]{taguchiea}
\bibinfo{author}{\bibfnamefont{Y.}~\bibnamefont{Taguchi}},
  \bibinfo{author}{\bibfnamefont{Y.}~\bibnamefont{Oohara}},
  \bibinfo{author}{\bibfnamefont{H.}~\bibnamefont{Yoshizawa}},
  \bibinfo{author}{\bibfnamefont{N.}~\bibnamefont{Nagaosa}}, \bibnamefont{and}
  \bibinfo{author}{\bibfnamefont{Y.}~\bibnamefont{Tokura}},
  \bibinfo{journal}{Science} \textbf{\bibinfo{volume}{291}},
  \bibinfo{eid}{2573} (\bibinfo{year}{2001}).

\bibitem[{\citenamefont{Jungwirth et~al.}(2002)\citenamefont{Jungwirth, Niu,
  and MacDonald}}]{jungwirthea}
\bibinfo{author}{\bibfnamefont{T.}~\bibnamefont{Jungwirth}},
  \bibinfo{author}{\bibfnamefont{Q.}~\bibnamefont{Niu}}, \bibnamefont{and}
  \bibinfo{author}{\bibfnamefont{A.~H.} \bibnamefont{MacDonald}},
  \bibinfo{journal}{Phys. Rev. Lett.} \textbf{\bibinfo{volume}{88}},
  \bibinfo{pages}{207208} (\bibinfo{year}{2002}).

\bibitem[{\citenamefont{Fang et~al.}(2003)\citenamefont{Fang, Nagaosa,
  Takahashi, Asamitsu, Mathieu, Ogasawara, Yamada, Masashi, Tokura, and
  Terakura}}]{fang}
\bibinfo{author}{\bibfnamefont{Z.}~\bibnamefont{Fang}},
  \bibinfo{author}{\bibfnamefont{N.}~\bibnamefont{Nagaosa}},
  \bibinfo{author}{\bibfnamefont{K.~S.} \bibnamefont{Takahashi}},
  \bibinfo{author}{\bibfnamefont{A.}~\bibnamefont{Asamitsu}},
  \bibinfo{author}{\bibfnamefont{R.}~\bibnamefont{Mathieu}},
  \bibinfo{author}{\bibfnamefont{T.}~\bibnamefont{Ogasawara}},
  \bibinfo{author}{\bibfnamefont{H.}~\bibnamefont{Yamada}},
  \bibinfo{author}{\bibnamefont{Masashi}},
  \bibinfo{author}{\bibfnamefont{Y.}~\bibnamefont{Tokura}}, \bibnamefont{and}
  \bibinfo{author}{\bibfnamefont{K.}~\bibnamefont{Terakura}},
  \bibinfo{journal}{Science} \textbf{\bibinfo{volume}{302}}, \bibinfo{eid}{92}
  (\bibinfo{year}{2003}).

\bibitem[{\citenamefont{Smit}(1958)}]{smit}
\bibinfo{author}{\bibfnamefont{J.}~\bibnamefont{Smit}},
  \bibinfo{journal}{Physica~(Amsterdam)} \textbf{\bibinfo{volume}{24}},
  \bibinfo{eid}{39} (\bibinfo{year}{1958}).

\bibitem[{\citenamefont{Berger}(1970)}]{berger}
\bibinfo{author}{\bibfnamefont{L.}~\bibnamefont{Berger}},
  \bibinfo{journal}{Phys. Rev. B} \textbf{\bibinfo{volume}{2}},
  \bibinfo{eid}{4559} (\bibinfo{year}{1970}).

\bibitem[{\citenamefont{Chien and Westgate}(1980)}]{chienwestgate}
\bibinfo{editor}{\bibfnamefont{C.~L.} \bibnamefont{Chien}} \bibnamefont{and}
  \bibinfo{editor}{\bibfnamefont{C.~R.} \bibnamefont{Westgate}}, eds.,
  \emph{\bibinfo{title}{The Hall Effect and Its Applications}}
  (\bibinfo{publisher}{Plenum Press,~New~York}, \bibinfo{year}{1980}).

\bibitem[{\citenamefont{Xiong et~al.}(1992)\citenamefont{Xiong, Xiao, Wang,
  Xiao, Jiang, and Chien}}]{xiongetal}
\bibinfo{author}{\bibfnamefont{P.}~\bibnamefont{Xiong}},
  \bibinfo{author}{\bibfnamefont{G.}~\bibnamefont{Xiao}},
  \bibinfo{author}{\bibfnamefont{J.~Q.} \bibnamefont{Wang}},
  \bibinfo{author}{\bibfnamefont{J.~Q.} \bibnamefont{Xiao}},
  \bibinfo{author}{\bibfnamefont{J.~S.} \bibnamefont{Jiang}}, \bibnamefont{and}
  \bibinfo{author}{\bibfnamefont{C.~L.} \bibnamefont{Chien}},
  \bibinfo{journal}{Phys. Rev. Lett.} \textbf{\bibinfo{volume}{69}},
  \bibinfo{eid}{3220} (\bibinfo{year}{1992}).

\bibitem[{\citenamefont{Xu et~al.}(2008)\citenamefont{Xu, Zhang, Wang, Chu, Li,
  Yu, and Zhang}}]{xuetal}
\bibinfo{author}{\bibfnamefont{W.~J.} \bibnamefont{Xu}},
  \bibinfo{author}{\bibfnamefont{B.}~\bibnamefont{Zhang}},
  \bibinfo{author}{\bibfnamefont{Z.}~\bibnamefont{Wang}},
  \bibinfo{author}{\bibfnamefont{S.}~\bibnamefont{Chu}},
  \bibinfo{author}{\bibfnamefont{W.}~\bibnamefont{Li}},
  \bibinfo{author}{\bibfnamefont{R.~H.} \bibnamefont{Yu}}, \bibnamefont{and}
  \bibinfo{author}{\bibfnamefont{X.~X.} \bibnamefont{Zhang}},
  \bibinfo{journal}{Eur. Phys. J. B}
  \textbf{\bibinfo{volume}{65}},
  \bibinfo{eid}{233} (\bibinfo{year}{2008}).

\bibitem[{\citenamefont{Langenfeld and W\"olfle}(1991)}]{langenfeldwolfle}
\bibinfo{author}{\bibfnamefont{A.}~\bibnamefont{Langenfeld}} \bibnamefont{and}
  \bibinfo{author}{\bibfnamefont{P.}~\bibnamefont{W\"olfle}},
  \bibinfo{journal}{Phys. Rev. Lett.} \textbf{\bibinfo{volume}{67}},
  \bibinfo{eid}{739} (\bibinfo{year}{1991}).

\bibitem[{\citenamefont{Muttalib and W\"olfle}(2007)}]{MW}
\bibinfo{author}{\bibfnamefont{K.~A.} \bibnamefont{Muttalib}} \bibnamefont{and}
  \bibinfo{author}{\bibfnamefont{P.}~\bibnamefont{W\"olfle}},
  \bibinfo{journal}{Phys. Rev. B} \textbf{\bibinfo{volume}{76}},
  \bibinfo{eid}{214415} (\bibinfo{year}{2007}).

\bibitem[{\citenamefont{Dugaev et~al.}(2001)\citenamefont{Dugaev, Cr\'epieux,
  and Bruno}}]{dugaevea}
\bibinfo{author}{\bibfnamefont{V.~K.} \bibnamefont{Dugaev}},
  \bibinfo{author}{\bibfnamefont{A.}~\bibnamefont{Cr\'epieux}},
  \bibnamefont{and} \bibinfo{author}{\bibfnamefont{P.}~\bibnamefont{Bruno}},
  \bibinfo{journal}{Phys. Rev. B} \textbf{\bibinfo{volume}{64}},
  \bibinfo{eid}{104411} (\bibinfo{year}{2001}).

\bibitem[{\citenamefont{Bergmann and Ye}(1991)}]{bergmannye}
\bibinfo{author}{\bibfnamefont{G.}~\bibnamefont{Bergmann}} \bibnamefont{and}
  \bibinfo{author}{\bibfnamefont{F.}~\bibnamefont{Ye}}, \bibinfo{journal}{Phys.
  Rev. Lett.} \textbf{\bibinfo{volume}{67}}, \bibinfo{eid}{735}
  (\bibinfo{year}{1991}).

\bibitem[{\citenamefont{Mitra et~al.}(2007)\citenamefont{Mitra, Misra, Hebard,
  Muttalib, and W\"olfle}}]{mitra}
\bibinfo{author}{\bibfnamefont{P.}~\bibnamefont{Mitra}},
  \bibinfo{author}{\bibfnamefont{R.}~\bibnamefont{Misra}},
  \bibinfo{author}{\bibfnamefont{A.~F.} \bibnamefont{Hebard}},
  \bibinfo{author}{\bibfnamefont{K.~A.} \bibnamefont{Muttalib}},
  \bibnamefont{and} \bibinfo{author}{\bibfnamefont{P.}~\bibnamefont{W\"olfle}},
  \bibinfo{journal}{Phys. Rev. Lett.} \textbf{\bibinfo{volume}{99}},
  \bibinfo{eid}{046804} (\bibinfo{year}{2007}).

\bibitem[{\citenamefont{Kharitonov and Efetov}(2007)}]{KEprl}
\bibinfo{author}{\bibfnamefont{M.~Y.} \bibnamefont{Kharitonov}}
  \bibnamefont{and} \bibinfo{author}{\bibfnamefont{K.~B.}
  \bibnamefont{Efetov}}, \bibinfo{journal}{Phys. Rev. Lett.}
  \textbf{\bibinfo{volume}{99}}, \bibinfo{eid}{056803} (\bibinfo{year}{2007}).

\bibitem[{\citenamefont{Kharitonov and Efetov}(2008)}]{KEprb}
\bibinfo{author}{\bibfnamefont{M.~Y.} \bibnamefont{Kharitonov}}
  \bibnamefont{and} \bibinfo{author}{\bibfnamefont{K.~B.}
  \bibnamefont{Efetov}}, \bibinfo{journal}{Phys. Rev. B}
  \textbf{\bibinfo{volume}{77}}, \bibinfo{eid}{045116} (\bibinfo{year}{2008}).

\bibitem[{\citenamefont{Beloborodov et~al.}(2007)\citenamefont{Beloborodov,
  Lopatin, Vinokur, and Efetov}}]{BLVErmp}
\bibinfo{author}{\bibfnamefont{I.~S.} \bibnamefont{Beloborodov}},
  \bibinfo{author}{\bibfnamefont{A.~V.} \bibnamefont{Lopatin}},
  \bibinfo{author}{\bibfnamefont{V.~M.} \bibnamefont{Vinokur}},
  \bibnamefont{and} \bibinfo{author}{\bibfnamefont{K.~B.}
  \bibnamefont{Efetov}}, \bibinfo{journal}{Rev. Mod. Phys.}
  \textbf{\bibinfo{volume}{79}}, \bibinfo{eid}{469} (\bibinfo{year}{2007}).

\bibitem[{\citenamefont{Abrikosov et~al.}(1965)\citenamefont{Abrikosov,
  Gor'kov, and Dzyaloshinski}}]{abrikosov}
\bibinfo{author}{\bibfnamefont{A.~A.} \bibnamefont{Abrikosov}},
  \bibinfo{author}{\bibfnamefont{L.~P.} \bibnamefont{Gor'kov}},
  \bibnamefont{and} \bibinfo{author}{\bibfnamefont{I.~E.}
  \bibnamefont{Dzyaloshinski}}, \emph{\bibinfo{title}{Methods of Quantum Field
  Theory in Statistical Physics}} (\bibinfo{publisher}{Dover,~New~York},
  \bibinfo{year}{1965}).

\bibitem[{\citenamefont{Cr\'epieux and Bruno}(2001)}]{crepieuxbruno}
\bibinfo{author}{\bibfnamefont{A.}~\bibnamefont{Cr\'epieux}} \bibnamefont{and}
  \bibinfo{author}{\bibfnamefont{P.}~\bibnamefont{Bruno}},
  \bibinfo{journal}{Phys. Rev. B} \textbf{\bibinfo{volume}{64}},
  \bibinfo{eid}{014416} (\bibinfo{year}{2001}).

\bibitem[{\citenamefont{Beloborodov et~al.}(2004)\citenamefont{Beloborodov,
  Lopatin, and Vinokur}}]{BLV}
\bibinfo{author}{\bibfnamefont{I.~S.} \bibnamefont{Beloborodov}},
  \bibinfo{author}{\bibfnamefont{A.~V.} \bibnamefont{Lopatin}},
  \bibnamefont{and} \bibinfo{author}{\bibfnamefont{V.~M.}
  \bibnamefont{Vinokur}}, \bibinfo{journal}{Phys. Rev. B}
  \textbf{\bibinfo{volume}{70}}, \bibinfo{pages}{205120}
  (\bibinfo{year}{2004}).

\bibitem[{\citenamefont{Biagini et~al.}(2005)\citenamefont{Biagini, Caneva,
  Tognetti, and Varlamov}}]{biagini}
\bibinfo{author}{\bibfnamefont{C.}~\bibnamefont{Biagini}},
  \bibinfo{author}{\bibfnamefont{T.}~\bibnamefont{Caneva}},
  \bibinfo{author}{\bibfnamefont{V.}~\bibnamefont{Tognetti}}, \bibnamefont{and}
  \bibinfo{author}{\bibfnamefont{A.~A.} \bibnamefont{Varlamov}},
  \bibinfo{journal}{Phys. Rev. B} \textbf{\bibinfo{volume}{72}},
  \bibinfo{eid}{041102} (\bibinfo{year}{2005}).

\bibitem[{\citenamefont{Beloborodov et~al.}(2001)\citenamefont{Beloborodov,
  Efetov, Altland, and Hekking}}]{BEAH}
\bibinfo{author}{\bibfnamefont{I.~S.} \bibnamefont{Beloborodov}},
  \bibinfo{author}{\bibfnamefont{K.~B.} \bibnamefont{Efetov}},
  \bibinfo{author}{\bibfnamefont{A.}~\bibnamefont{Altland}}, \bibnamefont{and}
  \bibinfo{author}{\bibfnamefont{F.~W.~J.} \bibnamefont{Hekking}},
  \bibinfo{journal}{Phys. Rev. B} \textbf{\bibinfo{volume}{63}},
  \bibinfo{pages}{115109} (\bibinfo{year}{2001}).

\bibitem[{\citenamefont{Vavilov and Glazman}(2003)}]{VG}
\bibinfo{author}{\bibfnamefont{M.}~\bibnamefont{Vavilov}} \bibnamefont{and}
  \bibinfo{author}{\bibfnamefont{L.}~\bibnamefont{Glazman}},
  \bibinfo{journal}{Phys. Rev. B} \textbf{\bibinfo{volume}{67}},
  \bibinfo{pages}{115310} (\bibinfo{year}{2003}).

\bibitem[{\citenamefont{Gor'kov et~al.}(1979)\citenamefont{Gor'kov, Larkin, and
  Khmel'nitski\u{\i}}}]{gorkovlarkinkhmelnitskii}
\bibinfo{author}{\bibfnamefont{L.~P.} \bibnamefont{Gor'kov}},
  \bibinfo{author}{\bibfnamefont{A.~I.} \bibnamefont{Larkin}},
  \bibnamefont{and} \bibinfo{author}{\bibfnamefont{D.~E.}
  \bibnamefont{Khmel'nitski\u{\i}}}, \bibinfo{journal}{Sov.~Phys. JETP Lett.}
  \textbf{\bibinfo{volume}{30}}, \bibinfo{pages}{228} (\bibinfo{year}{1979}).

\bibitem[{\citenamefont{Altshuler et~al.}(1982)\citenamefont{Altshuler, Aronov,
  and Khmel'nitzkii}}]{altshuleraronovkhmelnitsky}
\bibinfo{author}{\bibfnamefont{B.~L.} \bibnamefont{Altshuler}},
  \bibinfo{author}{\bibfnamefont{A.~G.} \bibnamefont{Aronov}},
  \bibnamefont{and} \bibinfo{author}{\bibfnamefont{D.~E.}
  \bibnamefont{Khmel'nitzkii}}, \bibinfo{journal}{J. Phys. C}
  \textbf{\bibinfo{volume}{15}}, \bibinfo{pages}{7367} (\bibinfo{year}{1982}).

\bibitem[{\citenamefont{Efetov and Tschersich}(2003)}]{ET}
\bibinfo{author}{\bibfnamefont{K.~B.} \bibnamefont{Efetov}} \bibnamefont{and}
  \bibinfo{author}{\bibfnamefont{A.}~\bibnamefont{Tschersich}},
  \bibinfo{journal}{Phys. Rev. B} \textbf{\bibinfo{volume}{67}},
  \bibinfo{pages}{174205} (\bibinfo{year}{2003}).

\bibitem[{\citenamefont{Beloborodov et~al.}(2003)\citenamefont{Beloborodov,
  Efetov, Lopatin, and Vinokur}}]{BELV}
\bibinfo{author}{\bibfnamefont{I.~S.} \bibnamefont{Beloborodov}},
  \bibinfo{author}{\bibfnamefont{K.~B.} \bibnamefont{Efetov}},
  \bibinfo{author}{\bibfnamefont{A.~V.} \bibnamefont{Lopatin}},
  \bibnamefont{and} \bibinfo{author}{\bibfnamefont{V.~M.}
  \bibnamefont{Vinokur}}, \bibinfo{journal}{Phys. Rev. Lett.}
  \textbf{\bibinfo{volume}{91}}, \bibinfo{pages}{246801}
  (\bibinfo{year}{2003}).

\end{thebibliography}
\end{document}